\documentclass[aps,pre,reprint, superscriptaddress]{revtex4-2}

\usepackage[utf8]{inputenc} % allow utf-8 input
\usepackage[T1]{fontenc}    % use 8-bit T1 fonts
\usepackage{lmodern}
\usepackage{amsfonts}       % blackboard math symbols
\usepackage{amsmath}
\usepackage{amsbsy}
\usepackage{float}

\usepackage{graphicx}
\graphicspath{{fig/}}

\usepackage{xcolor}
\usepackage{hyperref}

\begin{document}

\title{A simple model for conserved intracellular dynamics exhibits
multiscale pattern formation, traveling protein domains and arrested coarsening of lipid domains}

%\title{A simple model for conserved intracellular protein-lipid dynamics at membranes exhibits
%multiscale pattern formation, traveling domains and arrested coarsening}

\author{Benjamin Winkler}
\affiliation{
  Physikalisch-Technische Bundesanstalt, Abbestr. 2–12, 10587 Berlin, Germany
}
\affiliation{
  Freie Universität Berlin, Department of Mathematics and Computer Science, Arnimallee 3, 14195 Berlin, Germany
}

\author{Sergio Alonso}
\affiliation{
  Universitat Polit\`{e}cnica de Catalunya (UPC),
  Department of Physics,
  C. Jordi Girona, 1-3, 08034 Barcelona, Spain
}

\author{Markus Bär}
\affiliation{
  Physikalisch-Technische Bundesanstalt, Abbestr. 2–12, 10587 Berlin, Germany
}

\date{\today}

\begin{abstract}
\textbf{Abstract:} Herein, we model the spatiotemporal dynamics of cellular protein concentrations near membranes composed of different lipids by a three-variable continuum model describing the concentration of a protein attached to the membrane, the concentration of a protein freely diffusing in the cytosol  and the local  composition of lipids in a binary membrane. This model contains two globally conserved fields representing (i) the total protein content (sum of integrated cytosolic and membrane-bound protein concentrations) and (ii) the averaged fractions of the two lipid species in the membrane. The model combines a simple conserved reaction-diffusion model originally derived to study cell polarization via active phase separation (APS) for the protein dynamics and a Cahn-Hilliard equation describing phase separation or demixing of the two lipid species in the membrane. By means of linear stability analysis of the spatially homogeneous steady state of the systems and by direct numerical simulation of the full equations, we find that the lipid dynamics can undergo classical phase separation. 
In contrast, the protein dynamics exhibits an oscillatory phase separation for intermediate total protein contents related to a long-wavelength instability with complex eigenvalues and leading to traveling domains. 
Above a critical, mutual coupling and in the parameter regime of multiscale pattern formation, we find larger scale, traveling and rotating protein domains coexisting with smaller scale, stationary lipid domains exhibiting arrested coarsening.
It is also shown that the basic  instabilities and the phase diagram of the model are well captured by an extension of a recently proposed conserved FitzHugh-Nagumo model  for non-reciprocal pattern formation. This extended model is given by two non-reciprocally coupled Cahn-Hilliard equations allowing for instabilities of both involved species. 
This qualitative model  also contains substantially different interface tensions of the two considered species, mimicking the different physical properties of lipids and proteins in a cellular setting and  explaining the observed asymmetry in behavior (static lipid vs. traveling protein patterns), found in linear stability analysis and simulations of the full model.    
\end{abstract}

\maketitle

\section{Introduction}

The spatiotemporal organization of molecules is crucial for the biological regulation of living cells \cite{kholodenko2006cell}.
Self-assembly and self-organization are processes that govern the emergence of order from disordered components and are, therefore, key to understand the formation of complex structures in cell biology \cite{sackmann1995physical,nicolson2014fluid}. 
On the one hand, self-assembly determines the formation of structures in a process towards equilibrium by minimizing the free energy of the system.
Examples of such dynamics are the unassisted protein folding process \cite{dobson2004principles} and the phase separation in mixtures of two or more components \cite{cahn1958free}. 
On the other hand, self-organization needs the perpetual consumption of energy to maintain a certain dynamics, giving rise to waves or distinct domains \cite{beta2017intracellular}, e.g., underlying the emergent dynamics of the cytoskeleton~\cite{bretschneider2009three}. 
Both processes are sometimes confused and actually self-organization is also known as dynamic self-assembly \cite{whitesides2002self} or chemically fueled self-assembly \cite{das2021chemically}.

Lipid microdomains are formed in membranes of living cells \cite{edidin1997lipid,cebecauer2018membrane}. 
The organization of the different types of phospholipids at the cell membrane has been frequently related to the process of phase separation~\cite{sackmann1995physical}, because similar separation dynamics have been observed in giant vesicles composed of ternary mixtures of phospholipids and cholesterol \cite{veatch2003separation}. 
This relation has contributed to the hypothesis of \emph{lipid rafts}, which postulates that lipid–lipid interactions can organize cell membranes into domains of distinct structures with a size of about 100~$nm$ \cite{cebecauer2018membrane}, which have important roles for cellular function \cite{levental2020lipid,case2019regulation}. 
Lipid domains in membranes were also shown to be controlled or induced by interactions with proteins \cite{dietrich2009interaction} and enzymes such as kinases and phosphatases \cite{hansen2019stochastic}.

Complementary, the separation of the protein concentrations at the cell membrane can give rise to the formation of intracellular patterns (ICPs) and induce the polarization of living eukaryote cells~\cite{rappel2017mechanisms}.
Thereby, specific regions of the cell membrane are covered with certain proteins leading to a morphological and functional distinction of front and  back in the cell \cite{drubin1996origins,mogilner2012cell}. 
Such a process is typically controlled by enzymatic reactions and, therefore, actively maintained by the consumption of chemical energy in the form of ATP. 
Due to the typical cell size,  this process  of self-organization is tuned to  the characteristic length of around 10 $\mu m$.

Hence, while phospholipids generate domains by lipid-lipid interactions, different types of  proteins  simultaneously segregate at the cellular membrane \cite{saha2022active,levental2023regulation}.  
The specific interactions between these proteins and lipid nano-domains are mediated, e.g., by electrostatic interactions between lipid headgroups and specific amino acid sequences, as well as by differences in the conformational order of both the lipid assemblies and the proteins. 
Furthermore, it is experimentally known that proteins interact with the domains of lipids and can induce large scale fluid/fluid phase coexistence in systems with realistic biological membrane compositions \cite{baumgart2007large}, and that the binding of certain proteins can induce the formation of a textured lipid phase, giving rise to lipid rafts \cite{raghunathan2018dynamic}.

Different models addressing the formation of ICPs were created in the framework of continuous reaction-diffusion models, governed by systems of coupled nonlinear partial differential equations. 
The simplest non-equilibrium models consider one protein species that can be bound to the membrane and released into the cytosol. 
Such models usually assume that the total protein mass in the model is conserved. Examples include models for cell polarization triggered by phase separation of small Rho GTPases~\cite{mori2008wave}  as well as more specific models for polarization in Dictyostelium cells~\cite{otsuji2007mass}, in budding yeast cells~\cite{goryachev2008dynamics} or of the PAR protein distribution in C. Elegans~\cite{trong2014parameter}. 
A review article compares these systems and highlights the role of mass conservation \cite{halatek2018self}. 
Furthermore, a model of the myristoelectrostatic switch of MARCKS proteins near membranes of homogeneous composition considers a  membrane bound protein as well as two cytosolic forms leading to self-organisation (or active phase separation) of the membrane~\cite{john2005alternative} and to bistable dynamics~\cite{alonso2010phase}. 
The similarities of many of these processes have also inspired a number of more general mathematical and theoretical analysis of such phenomena \cite{morita2010stability,mori2011asymptotic,halatek2018rethinking,tateno2021interfacial}.
Near the typical long-wavelength instability (type II-S in the classification of Cross and Hohenberg \cite{cross1993pattern}) in all the non-equilibrium reaction-diffusion systems mentioned above, the Cahn-Hilliard equation, 
originally derived as a continuum model for passive phase separation~\cite{cahn1958free}, was found to be the universal amplitude equation~\cite{bergmann2018active,bergmann2019system}. 
Such a derivation was also shown to be universally valid \cite{bergmann2018active} for continuum models exhibiting type II-S instabilities, such as motility induced phase separation  of active Brownian particles \cite{cates2015motility} or mechanochemical patterns in active viscous fluids \cite{bois2011pattern}.  
It is worth noting, that these systems show mostly coarsening domains or labyrinthine patterns. 
If the conservation of the considered concentrations is violated, e.g., by allowing exchange of material or chemical reactions, a characteristic wavelength is selected and the coarsening process is arrested~\cite{glotzer1995reaction,brauns2021wavelength}. 

The most prominent example for pattern formation in cellular systems is probably the so-called Min protein dynamics.
It describes the dynamics of the MinD and Min E proteins which are both present in various cytosolic and membrane bound forms of the bacterium E. Coli.
Hence, they are modeled by coupled reaction-diffusion equations that contain conserved quantities, namely the total concentration of MinD and MinE~\cite{kessler2016nonlinear,wettmann2018min,halatek2018self}. 
In experiments, standing and traveling waves in the Min system were observed in-vivo \cite{raskin1999minde} and in-vitro~\cite{loose2008spatial}. 
In-vivo, the standing waves are most common and control the mid point selection before cell division in E. Coli cells~\cite{lutkenhaus2007assembly}. 
More recently, the interaction of \mbox{Rho GTPase} activity and F-actin polymerization in Xenopus and starfish oocytes have also shown an interesting variety of 
wave patterns and active turbulence in experiments and simulations of the corresponding model equations \cite{bement2015activator,wigbers2021hierarchy,michaud2022versatile,bement2024patterning}. 
Observations of protein waves visible in cellular phenomena, such as circular dorsal ruffles, were also successfully modeled by 
reaction-diffusion type model equations \cite{bernitt2017fronts}. 
Many of these models combine conserved quantities (typically total protein concentrations) with non-conserved variables (e.g. actin), for a recent review see~\cite{beta2023actin}. 
Mechanochemical pattern formation in viscous fluids with two regulating species \cite{kumar2014pulsatory} and in viscoelastic solids and fluids \cite{radszuweit2013intracellular,alonso2017mechanochemical} also yields waves and oscillations. 
A simple model that leads to traveling waves and oscillatory long-wavelength instability (type II-0 according to the classification of
Cross-Hohenberg) is given by non-reciprocally coupled Cahn-Hilliard-equations describing the generic nonequilbrium dynamics of two coupled 
conserved fields~\cite{saha2020scalar,you2020nonreciprocity,frohoff2021localized}.
It has yielded a large variety of complex patterns in recent studies see, e.g.,~\cite{saha2025effervescence,rana2024defect,parkavousi2024enhanced,brauns2024nonreciprocal,greve2025coexistence}. 
While this model was originally derived for binary mixtures with non-reciprocal mechanical interactions between the involved species \cite{saha2020scalar} and has exhibited arrested coarsening and simple traveling waves,
recent work has shown that these equations also provide the universal amplitude equations describing the dynamics for systems near a type II-0 instability 
\cite{frohoff2023nonreciprocal,greve2024amplitude}, thereby adding a missing piece to the general theory of pattern formation. 
In \cite{greve2024amplitude} it is shown analytically that arrested coarsening should generically occur for oscillatory phase separation (type II-0 instability), shedding light on and potentially explaining numerous numerical findings of arrested coarsening in model systems. 

The model studied in this paper is motivated by the observation that coupling active phase separation—arising from the cycling of a single protein at the membrane—to passive lipid phase separation can lead to oscillatory protein–lipid phase separation in the form of traveling domains, as observed for MARCKS proteins in the myristoyl-electrostatic switch.
In the detailed model of this process \cite{john2005travelling}, these larger scale traveling domains were attributed to an active process and well separated from small scale lipid-lipid phase separation in the membrane. 
Here, we study a simpler version of this original model and explore the case where the two instabilities of an active and a passive phase separation occur simultaneously at quite different length scales.  
The simultaneous occurrence of such instabilities often leads to new phenomena and was studied before in the vicinity of the corresponding co-dimension-2 bifurcation, e.g., the Turing-Hopf dynamics \cite{de1996spatiotemporal,de2007spatial}, the Turing-Wave dynamics \cite{yang2002pattern,nicola2002drifting,schuler2014spatio} or the non-reciprocal coupling of two non-conserved quantities described by coupled Swift-Hohenberg equations \cite{tateyama2024pattern,tateyama2026higher}.
Experiments and modeling in chemical pattern forming systems like the BZ-AOT systems showed that for largely different diffusion constants, and correspondingly largely different instability wavelengths, interesting and complex multiscale pattern formation occurs~\cite{vanag2001pattern,epstein2005complex,alonso2011complex}.
In this paper, we will explore the ramifications of simultaneously occurring phase separation processes with largely different length scales, i.e., active phase separation of proteins coupled with passive separation of lipids in membranes.

The paper is structured as follows: 
In Section II, we will introduce a simple model containing two conserved fields describing first, the total concentration of proteins - consisting of membrane-bound and cytosolic molecules - and second, the lipid composition of a two-component membrane. 
Section III contains the results of our study of the resulting model equations. 
In Section III.A, we vary the total mass of proteins and the lipid composition of the membrane as control parameters and carry out a linear stability analysis of the spatially homogeneous steady state of the model, yielding a characteristic phase diagram. 
In Section III.B, the results of numerical simulations of the nonlinear dynamics of the model are given. 
In Sections III.C and III.D, we focus on the regime where both, lipids and proteins exhibit phase separation instabilities and multiscale patterns are found for sufficiently strong coupling. 
Section III.C provides a detailed account of the emergence of traveling protein domains and the dependency of the domain velocity on the coupling strength between the conserved fields.
Section III.D reports on the evolution of the characteristic lipid domain scales on this coupling strength. 
Section III.E provides a minimal model of two non-reciprocally coupled Cahn-Hilliard equations that have the same characteristic instabilities as the full model explaining crucial aspects of the behavior close to the respective instabilities.
In Section IV, we conclude by summarizing the main findings and discussing them in the broader cell-biological and theoretical context of emergent complexity in biological systems.

\section{Model for Cell Polarization}

\begin{figure}[h!]
    \centering
    \includegraphics[width=\linewidth]{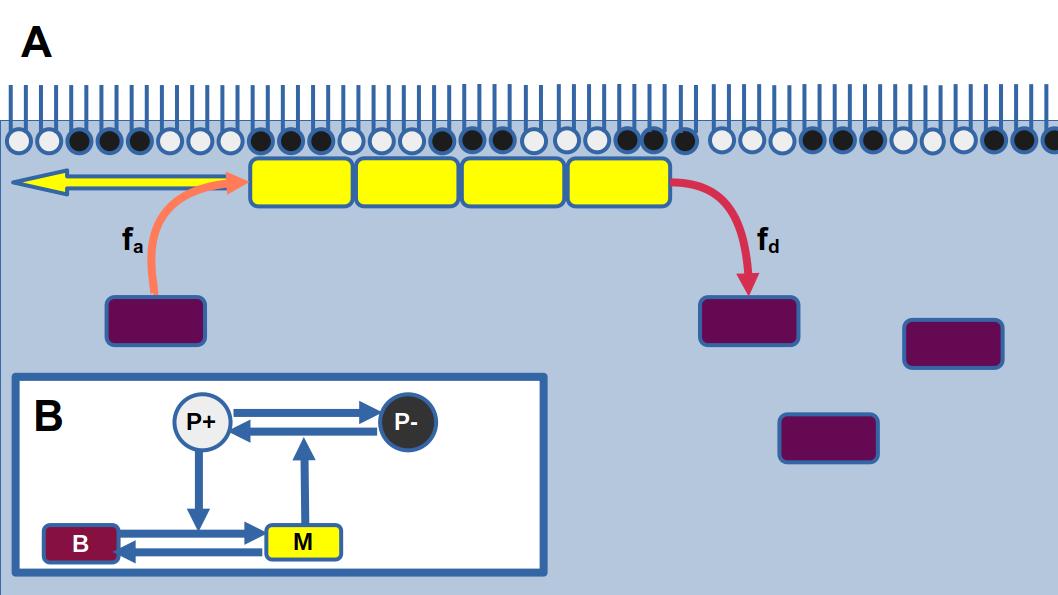}

    \caption{ Schematic of the model mechanics. A) The cell membrane's two species of lipid acids (black and white head groups) undergo coarsening.  The protein subsystem is composed out of a protein in its active and inactive form. The inactive, fast diffusing species in bulk (labeled B, in lilac) binds with the rate $f_a$ to the cell membrane and becomes active  and much less mobile (labeled M, in yellow). A dissociation rate $f_b$, linear in the amount of bound protein, recycles the inactive form. B) Both systems are coupled by first, $f_a$ and $f_b$ dependending on the local lipid composition and second, a shift in the chemical potential of the lipid-lipid interaction that depends on the local concentration of bound protein, cf. equations~1-3. The resulting protein pattern, a spot with higher M concentration that is stationary in the uncoupled case,  travels with a velocity depending on the coupling strength of the two systems.}
    \label{fig:model}
\end{figure}

Modeling on many different levels of complexity and specificity is an innovative tool of knowledge generation in cell biology and plays a pivotal role, e.g., in the investigation of mechanisms for cell polarity emergence~\cite{mogilner2012cell}.
Actively-driven, mass-conserved reaction diffusion systems, describing active phase separation (APS),  were proposed to describe the emergence of cell polarity~\cite{otsuji2007mass,mori2008wave,morita2010stability,bergmann2019system}.
We consider here a minimal model  of APS, featuring a pair of conserved protein species interacting with the cell membrane.
It leads to persistent cell polarity governed by one side of the cell with a phase that exhibits a high concentration of membrane-bound protein~\cite{mori2008wave,mori2011asymptotic}, cf. figure~\ref{fig:model}.
The pair of proteins represents a small Rho-GTPase in its inactive, quickly-diffusing form (here called $B$ for bulk) and $M$ the  membrane-bound, active form.
Due to the binding to the cell membrane, the diffusion of species $M$ is much slower:
%- for numerical purposes, we set here 
%$D_M=5 \ll D_B=200$ 
$D_M=0.25 \mu m^2 /s \ll D_B=10 \mu m ^2 /s$  \cite{luby1999cytoarchitecture}
%and discuss the conversion into physical units at the end of the section.
The ratio of both species is determined by the  difference between the rate of association to the membrane $f_a=B \left(\frac{M^2}{1+M^2}+k_0 \right)$ and a linear rate of dissociation $f_d=M$.
The association rate is proportional to the bulk concentration and consists of a small constant binding rate $k_0=0.067 s^{-1}$ and a nonlinear term modeling the positive feedback on recruitment of the protein at the membrane, i.e., the $M$ species, due to a self-attraction mediated by electrostatic interactions.
The APS system is then governed by a pair of coupled equations, $\partial_t M = D_M \Delta M + f(M,B),\,\, \partial_t B = D_B \Delta B - f(M,B)$, with the total amount of protein $E=M+B$ conserved, cf. equations (\ref{eq_sys1}) and (\ref{eq_sys2}).
For a certain amount of protein $E_0=M_0+B_0$ as homogeneous initial condition, the M-distribution separates into two distinct phases, whereas the bulk species $B$ remains roughly homogeneous.
\newline

The cell membrane, i.e.,  the substrate for the APS system's binding and unbinding, we consider as a demixing emulsion of two separate species of lipids, governed by Cahn-Hilliard dynamics, cf. equation~(\ref{eq_sys3}).
The difference between the lipid species in the cell can, e.g., stem from varied head groups with different electrostatic charges.
The term $\propto p^2$ expresses a difference in the free energy of both phases which shifts the phase-coexistence region towards positive values of the averaged, conserved $p_0$.
Hence, at $p_0=0$ we see droplets of the $p=+1$ phase in the background of the $p=-1$ phase, instead of symmetric labyrinthine patterns when the free energy of both phases is equal.
The phases p=+1 and p=-1 can be associated with two different lipid species making up the membrane of the cells \cite{veatch2003separation,veatch2005miscibility,zhao2007phase}. 
Here, we set $\tau=0.4$, 
%$D_p=0.5$ and $\Psi=0.1$ 
$D_p=0.025 \mu m^2/s$ and $\Psi=0.005 \mu m^2 $ 
which leads to significantly smaller droplet size than the protein pattern (although slowly coarsening with time), cf. figure~\ref{fig:simulation}.
With that, we arrive at the complete  system equations:

\begin{align}
\label{eq_sys1}
\partial_t B &= D_B \Delta B - (1+\alpha p) B \left(\frac{M^2}{1+M^2}+k_0 \right) + M\\
\label{eq_sys2}
\partial_t M &= D_M \Delta M + (1+\alpha p) B \left(\frac{M^2}{1+M^2}+k_0 \right) -  M\\
\label{eq_sys3}
\partial_t p &= D_p \Delta (p^3 - p^2-\tau p+\beta M - \Psi \Delta p)
\end{align}

The graphical depiction of the reaction equilibria of the two uncoupled subsystems is given in the supplementary materials as SM1.
As the binding affinities of the proteins with the membrane depend on the lipid species present in a real cell membrane, a spatial heterogeneity of the $p$ will lead to different binding rates with the $M-B$ subsystem. 
We capture this by coupling the two subsystems, i.e., the coupling parameter $\alpha \in [0,1]$ leads to a preferred binding of the $M$ species to the + phase rather than the - phase of the membrane.
More concretely, we modified the association rate by a factor of $(1+\alpha p)$ to capture the first order of such a variation in binding strength.
The association of the protein to the membrane generally also alters the local lipid environment, e.g., due to changes in the order of the lipid tails via insertion and the introduction or screening of electrostatic interactions of the head groups.
The modulation of the chemical potential due to the presence of bound protein is introduced by the term $\beta M$ in the Cahn-Hillard dynamics with the coupling parameter  $\beta \in [0,1]$.
For convenience we introduce the \emph{coupling strength} $\epsilon=\sqrt{\alpha\beta}$ and refer typically to the symmetric case $\alpha=\beta=\epsilon$ throughout the manuscript.
\newline

\section{Results}

\subsection{Phase Diagram: Linear Stability Analysis}\label{LSA}

\begin{figure*}[t!]
  \centering
  \includegraphics[width=\linewidth]{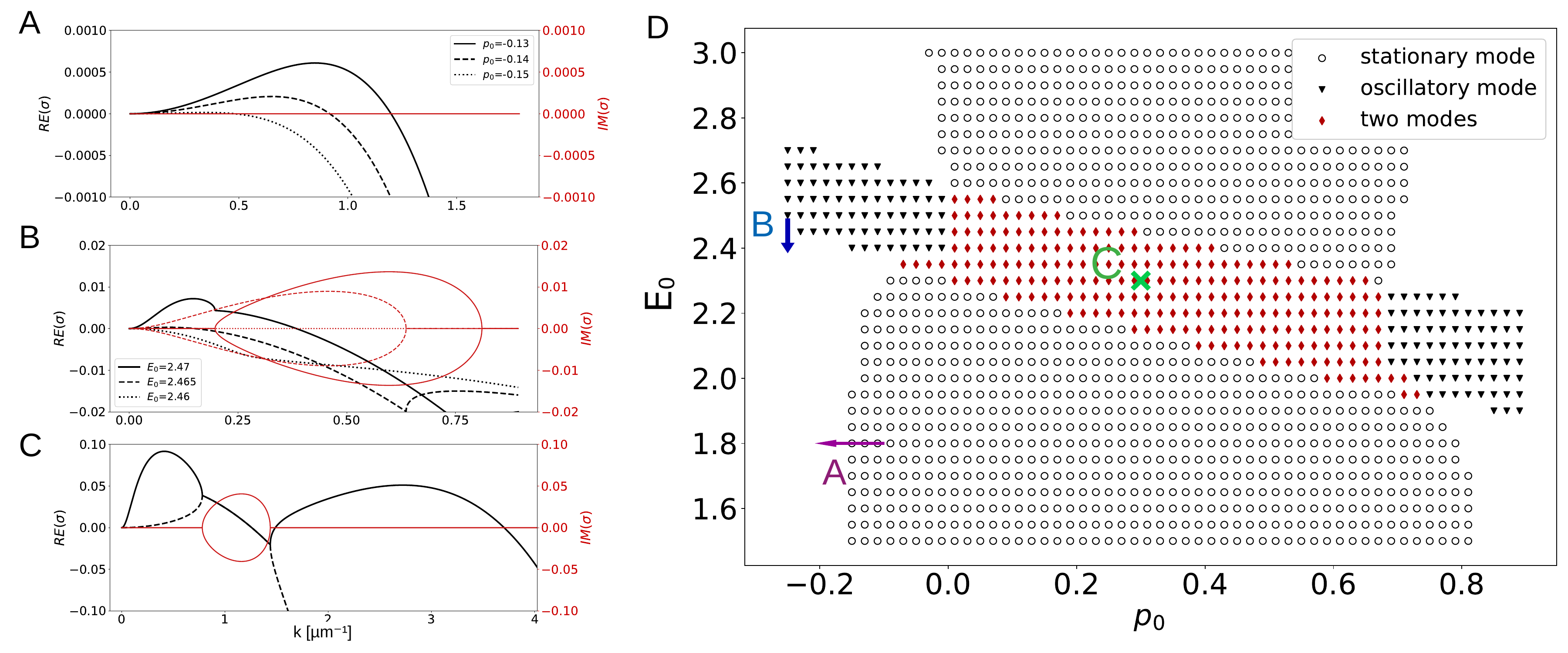}
    \caption{ Linear stability analysis of the coupled system with $\alpha=\beta=0.3$. A-C: Dispersion relations from the linear stability analysis of well-mixed initial conditions with different, homogeneous concentrations $(p_0, E_0)$.     
    A: The stationary instability, mediated by a change in $p_0$ with $E_0=B_0+M_0=1.8$ remaining constant. It is associated with the emergence of the  Cahn-Hilliard pattern, roughly existing for $p_0 \in [-0.15, 0.85]$. The imaginary part remains zero, whereas the fastest growing mode shifts towards larger k values towards the middle of the instability region, indicated by the positive maximum of the real part at $k\approx2.7 \mu m ^{-1}$. B: An oscillatory instability, associated with the emergence of the APS pattern, mediated by changes in the protein concentration while $p_0=-0.25$ remains constant. The associated FGM remains small, approximately $0.45 \mu m^{-1}$. C: In the middle of the instability region at $(p_0, E_0)=(0.3,2.3)$, both pattern are stable which is indicated by two separate, positive maxima in the dispersion relation. Note that the nonzero imaginary part does not occur at the fastest growing modes, but rather in between the two maxima. There, in the region framed by the two kinks in the curve, the real parts of two eigenvalues fall together.
     D) The phase diagram obtained from the linear stability analysis by classifying the dispersion relations with respect to the number of occurring maxima in the dispersion relation with $Re>0$ and the presence of oscillatory modes. The empty circles represent the membrane ($p_0$) and protein compositions ($E_0=M_0+B_0$) where only the stationary Cahn-Hilliard pattern forms, filled triangles stand for the existence of only the oscillatory APS pattern and both pattern existing simultaneously is indicated by filled, red diamonds. The transitions for changing either the membrane composition $p_0$ (indicated by A) or the total protein concentration (indicated by B), as well as the exemplary case C correspond to the dispersion relations shown in panels A-C. The supplementary material contains videos with  the systematic change of the dispersion curves in the transitions shown here (SM2 -  SM5) .}
    \label{fig:linstab}
\end{figure*}

We perform the linear stability analysis considering the Jacobi matrix of partial derivatives of the linearized system around a homogeneous steady state given by $(p_0,M_0,B_0)$ and a coupling of $\alpha = \beta =0.3$.
The resulting dispersion relations, cf. figure~\ref{fig:linstab} panels A, B and C, were calculated systematically for various initial values of $E_0=B_0+M_0$ and $p_0$ and classified with respect to the existence of unstable modes, i.e.,  positive maxima of the real part of the complex eigenvalue $\sigma$, and whether unstable modes with $Im(\sigma) \neq 0$ exist.
The resulting phase diagram mirrors that of the explicit system simulations, cf. figure~\ref{fig:simulation}.
\newline

First, we line out briefly what states we address in the analysis.
For the membrane system, each homogeneous state $p(x,y)=p_0$ is a steady state, as $\partial_t p_0 = 0$ holds, following Eq.~(\ref{eq_sys3}).
For the protein species, the possible homogeneous steady states are given by their reactive equilibrium $f(M_0,B_0,p_0)=0$ as then also $\partial_t M_0 = 0$ and $\partial_t B_0 = 0$ hold.
Hence, we fix $B_0$ and $p_0$ to calculate $M_0$ and thereby define the homogeneous steady states of different protein and lipid content in the systems as pairs $(p_0,E_0)$.

\begin{figure*}[t!]
    \centering
    \includegraphics[width=0.90\linewidth]{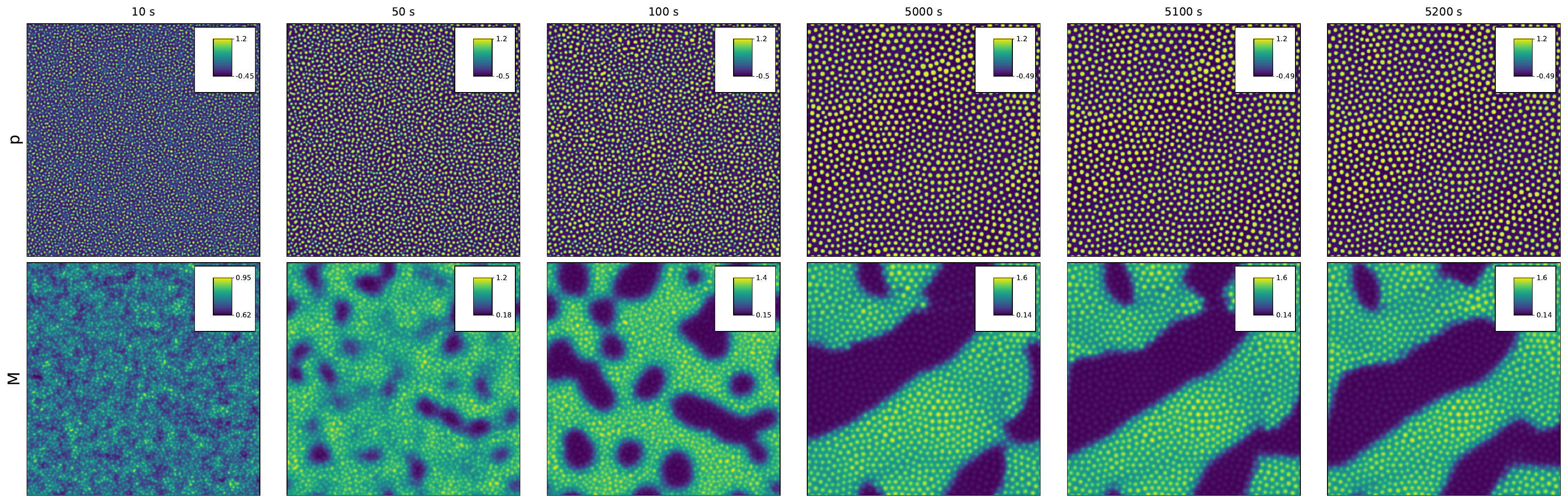}
    \caption{The time evolution of the patterns in a simulation of the coupled system $(\alpha=\beta=0.3)$ of size $33.5 \times 33.5 \mu m$ with periodic boundary conditions in 2D. The initial lipid composition is $p_0=0$ and the protein content $E_0=M_0+B_0=2.5$. In the upper row, the developing lipid pattern is shown whereas the formation of the larger-scale protein pattern is depicted in the lower row for $t=10s, 50s, 100s, 5000s, 5100s$ and $5200s$, respectively. Dynamic color ranges are given in the small insets in the upper right corners.
    The lipid pattern forms as very small spots - already at $t=10s$ - and slowly coarsens with time until it stops due to the coupling to the protein subsystem. The large pattern of the membrane-bound protein takes longer to form (growing spots at 50s to 100s) and starts to move subsequently. Note that due to the coupling the smaller lipid pattern also induces the same variations in the protein pattern and vice versa. This becomes especially apparent in the three right-most frames where the moving fronts of the protein pattern are also visible in the otherwise very regular lipid spot pattern.}
    \label{fig:patDevelopment}
\end{figure*}

In the resulting dispersion relations of the three eigenvalues $EV(k)$ we identify four cases: 
First, the  real value of all eigenvalues is smaller than zero for all wavenumbers $k$, hence, the homogeneous state given by $(p_0,M_0,B_0,\alpha,\beta)$ is linearly stable.
Second, a stationary instability occurs where the fastest growing mode (FGM) quickly moves to higher wavenumbers $k_{FGM}\approx 2.7 \mu m^{-1}$ - this is associated with the smaller Cahn-Hilliard pattern of the membrane, cf. figure~\ref{fig:linstab}, in panel A. 
This transition skirts the roughly vertical region of $p_0\in [-0.15,0.8]$ in figure~\ref{fig:linstab} panel D - indicated by open circles.
Third, an oscillatory instability occurs where the FGM grows much less, only until $k_{FGM}\approx 0.45 \mu m^{-1}$. 
This corresponds to the case of the bigger APS pattern being stable, cf. figure~\ref{fig:linstab} panel B.
This instability represents the borders for the stripe of filled triangles in the phase diagram for $E_0 \approx 2.6$ at $p_0=-0.3$ and decreasing to $E_0 \approx 2.1$ at $p_0=0.9$ due to the coupling to the p subsystem.
In this region we observe a single peak of positive eigenvalues at low k-values  - typically not the FGM, respectively - with a nonzero imaginary part.
And lastly, we observe a region where both stripes in panel D cross and two maxima with positive real part exist.
Hence, both patterns exist for the tuple of system parameters, cf.  figure~\ref{fig:linstab} panel C for the typical case.
There are always three eigenvalues, but for clarity we show here only the maximum of the real parts for each k-value. 
One of the eigenvalues is always negative and monotonously falling with k.
The two remaining eigenvalues can each exhibit one maximum that - depending on system parameters - may become positive, i.e., unstable.
In the intermediate region between the peaks, limited by two wavenumbers $k_1$ and $k_2>k_1$ that generally depend on system parameters, the two eigenvalues split up into a pair of conjugate-complex eigenvalues.
In figure~\ref{fig:linstab} panel C, this is shown for $(p_0,E_0)=(0.3, 2.3)$
At $k_1$ and $k_2$, respectively, there are kinks in the curve of the real parts, separating it from the regions with differing real part, i.e., $k<k_1$ and $k>k_2$.
Animated versions of the transition in terms of the maximum real eigenvalue can be found in the supplementary materials, i.e., (SM2 - SM5).

\subsection{Phase Diagram: Numerical Simulations}

\begin{figure*}[t!]
    \includegraphics[width=0.7\textwidth]{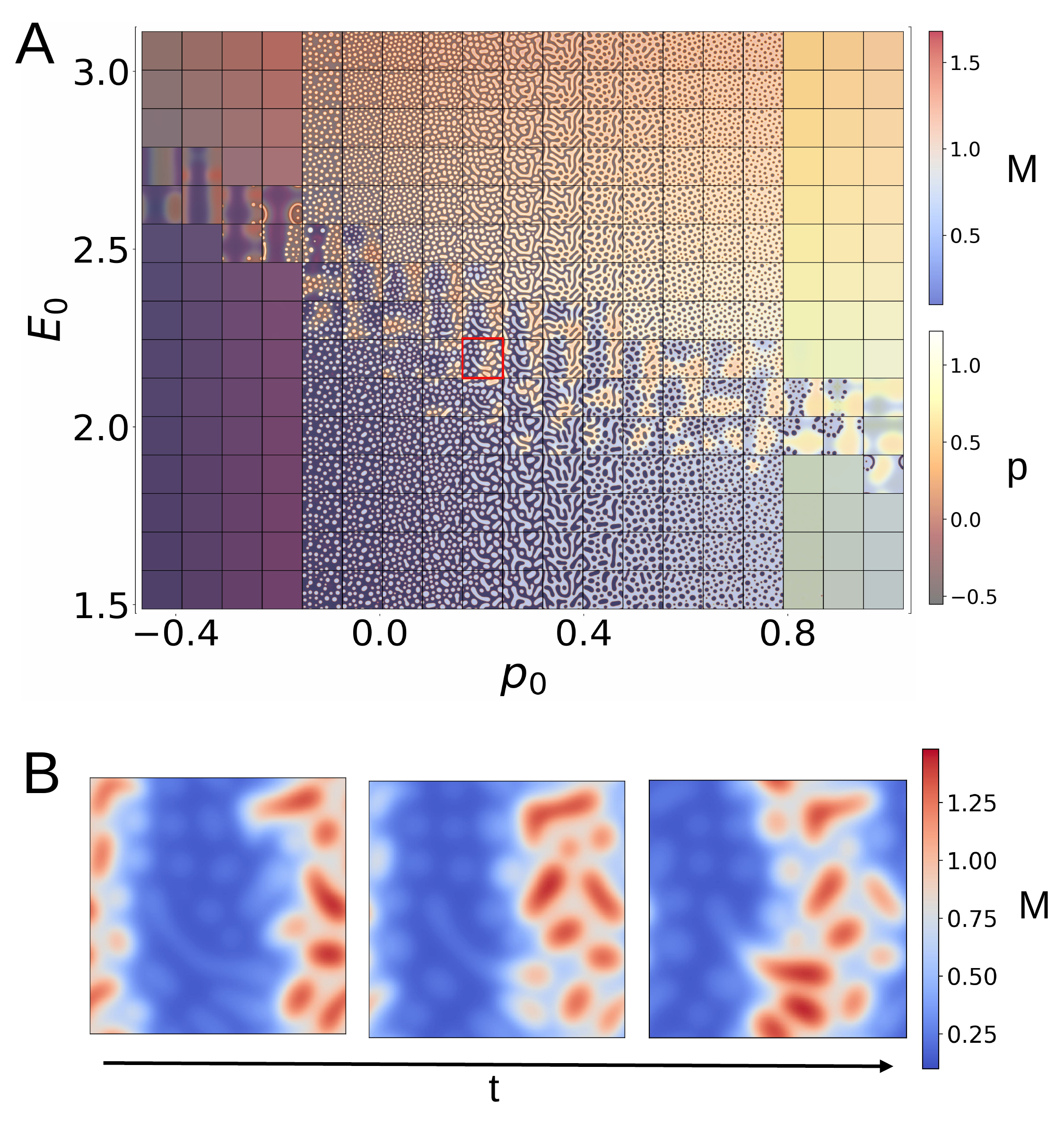}
    \caption{Phase diagram of the coupled system for $(\alpha=\beta=0.3)$ of size $16.8 \mu m$ with periodic boundary conditions in 2D.  A: For different initial conditions of the well-mixed system we obtain a phase diagram  for different protein content $E_0=(M_0+B_0) \in [1.5,3.1]$ in increments of $0.1$ and different lipid composition $p_0 \in [-0.5,0.9]$ in increments of $0.1$. The membrane ($p$ as a color gradient from dark gray over orange and yellow to white) demixes into small droplets for $-0.1<p<0.8$. The membrane-bound protein $M$  exists in a homogeneous state of either low or high concentration (blue over white to red overlaid). It forms the bigger APS pattern only in a small region of protein content falling from $E_0\approx 2.6$ to $E_0\approx 2$ over the shown range of lipid composition. Where both pattern coexist, the APS pattern exhibits a finite velocity. B: The time series  corresponds to three snapshots of the M species for the state framed in red with $(p_0,M_0+B_0)=(0.2,2.2)$. It wanders as a single front from right to left. Note that due to the coupling to the p-subsystem also the smaller pattern of the demixing membrane is visible as spots.}
    \label{fig:simulation}
\end{figure*}

After we have identified the regions in the phase space where spatial or spatiotemporal patterns can be expected as a result of the instability
of the homogeneous steady state, the corresponding numerical simulations of the full Eqs. (1-3) yield the nonlinear pattern dynamics, cf. figure~\ref{fig:patDevelopment}. 
Depicted there are the fast-developing, smaller-scale patterns of the lipids (p, top row) and the slower and larger-scale pattern of the membrane-bound proteins (M, bottom row).
The mutual coupling also makes the influence of each pattern on the other clearly visible, i.e., the small spots of the p-pattern in the M-field, as well as the domain boundaries of the M-pattern in the regular arrangement of the p-spots.

To generate an overview of the different qualitativ model pattern dynamics, we solved  Eqs. (1-3) numerically for the same parameters as used in the
linear stability analysis and a number of different initial conditions.
For the Cahn-Hilliard subsystem a homogeneous lipid composition $p_0\in[-0.5,1]$ was chosen.
Independently, the local protein content was set to an initially constant value of $E_0=M_0+B_0\in[1.5,3.1]$ with a symmetric split between $B_0$ and $M_0$ and zero-mean noise. 
The pattern forming M-distribution, together with the p-distribution of the membrane, is depicted as  snapshots of systems with size $L=33.5 \mu m$ at $t=500 s$ in figure~\ref{fig:simulation}, panel A.
The distributions of both systems are overlaid with different color schemes.
The p-distribution is shown as a gradient from black at the minimum value $p_{min} \approx -0.55$, over red at
$p \approx 0$ to  yellow and white, where p is maximal in the parameter range with $p_{max} \approx 1.2$.
The M-concentration as a gradient from blue to red  in the respective range $M \in [0.1,1.65]$.

The phase diagram is divided vertically into a region of low (roughly $E_0<2$) and a region of  high bound protein concentration M (for $E_0>2$) with a narrow transition zone.
Note that the chosen $E_0$ increases in the vertical direction linearly, but the amount of bound protein (M species) increases nonlinearly and differs locally due to the self-attraction, cf. equation~(2).
The membrane-bound protein pattern exists only in a narrow region between high and low M-concentration of the phase diagram, corresponding to a multi-stability region of the reactive equilibrium of the homogeneous state, cf. SM1 of the supplementary material for a depiction of the reaction equillibria.
The variation of this transition zone with respect to the lipid composition (in the horizontal direction of the depicted phase diagram) is due to the coupling of the systems, i.e. the p-dependence of the association rate of the protein to the membrane.
Furthermore, although the presence of a full coupling with the p-subsystem is necessary for the M-pattern to be able to move, the presence of a coarsening or spot-like pattern in the membrane is not, i.e., we observe moving APS pattern for each initial p-value in the presented range.
Note that, due to the coupling, inhomogeneities in one sub-system are transferred to the other.
With increasing coupling strength, the thereby developing gradients interact and destabilize the static nature of the (uncoupled) pattern.

The pattern of the Cahn-Hilliard system is observed in the vertical region of roughly $p_0\in [-0.15,0.8]$ as small spots of the + phase in a background of - phase for low initial $p_0$ and vice-versa for high initial $p_0$.
Right in the middle of the phase separating region ($p_0\approx 0.3-0.4$), the emerging pattern are labyrinthian.
In a central region of the $(E_0,p_0)$ plane both patterns coexist.
Snapshots in panels B depict an example M-pattern moving with a constant velocity.
However, the regularity of the motion in 2D is only given for small systems where a single band of the M pattern emerges. 
The behavior in the regime of multiscale pattern formation is illustrated in more detail in the movie SM6 in supplementary materials,
where one can see that first, two quasi-steady pattern for lipid and proteins are forming. Subsequently, the coupling leads to moving protein domains
which in turn affect the shape and size of the lipid domains. 
For larger systems, the number of moving fronts increases, and they move in random directions with regular mergers and break-ups, cf. SM6 in the supplementary material.
Hence, for a more systematic treatment of the protein pattern velocity, we will consider 1D systems, cf. section~\ref{PPV}.

\subsection{Onset of Motion and Velocities of One-Dimensional Protein Domains in Multiscale Pattern Formation}\label{PPV}

So far, we have studied the effect of the lipid composition and total protein mass, on the emerging patterns and identified three different behaviors: 
static phase separation of the lipids, traveling domains of proteins as well as multiscale pattern formation involving coexisting lipid and protein patterns at very different length scale and of different qualitative nature (traveling vs. stationary domains). 
Now, we focus on the role of the coupling strength between lipids and proteins in the multiscale pattern regime by varying the parameters $\alpha, \beta$, or $\epsilon = \sqrt{\alpha \beta}$ as coupling strength.
To study the motion of the protein pattern in more detail, we chose to examine 1D systems with appropriate initial values, i.e.,  with a neutral lipid composition ($p_0=0$) and a medium amount of protein ($E_0=2.5$).
In 1D it is easier to determine the velocity of a single moving spot by calculating the center of mass and the number of times it circles the entire domain per (sufficiently large) $\Delta t$.
As we are especially interested in the coupling-induced phenomena of the two systems, we systematically calculated the M-pattern velocity with respect to different coupling parameters $\alpha$ and $\beta$.

\begin{figure*}[t!]
    \centering
    \includegraphics[width=0.8\linewidth]{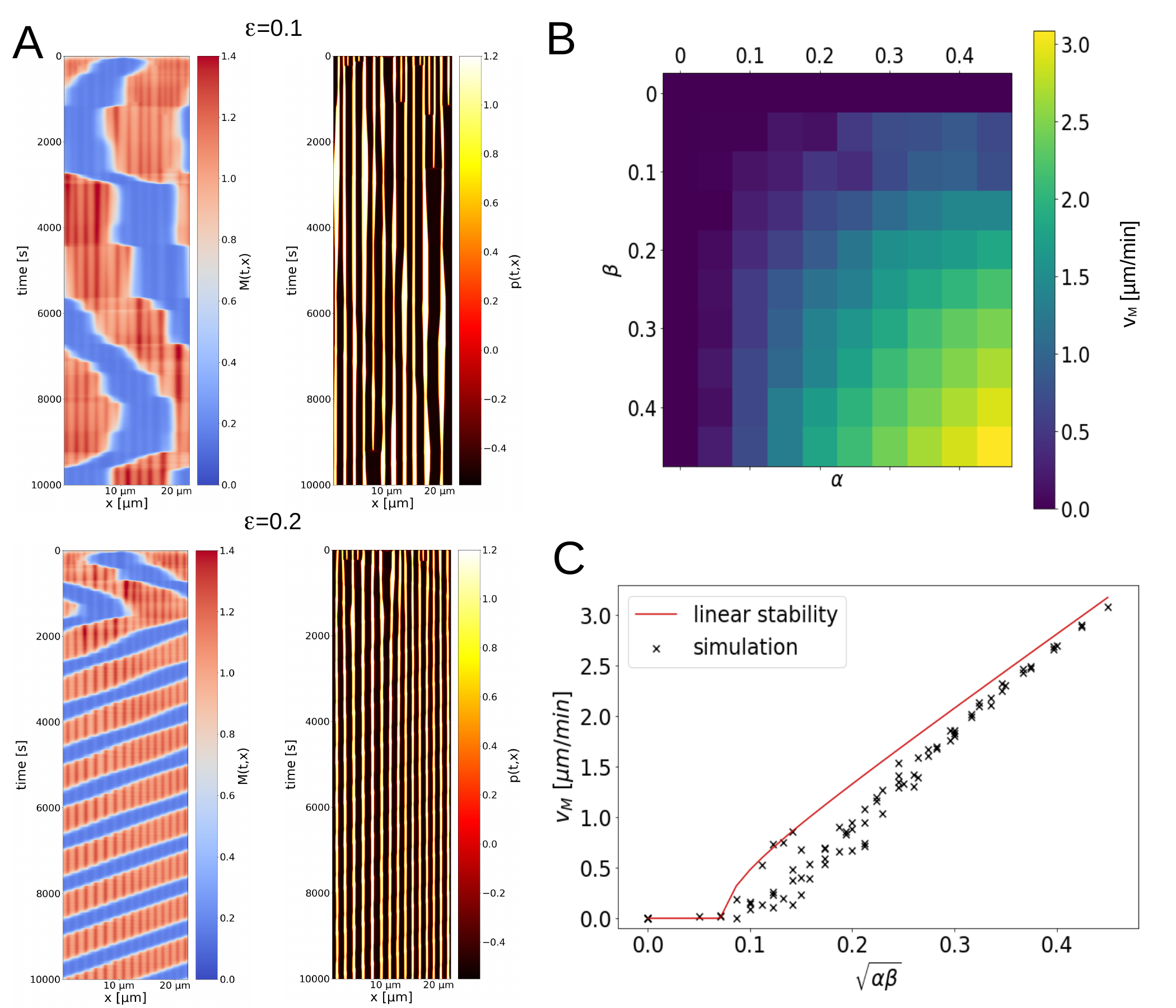}

    \caption{Protein pattern velocity investigated in 1D. A: The measured velocity of the M-pattern (left in blue to red color scheme) for low coupling $\epsilon=0.1$ is stochastic in nature, seemingly transiently pinned by the p-pattern (right). For larger couplings ($\epsilon=0.2$, below), the velocity becomes larger and more regular. Also the coarsening p-pattern is influenced by the M-distribution - small deformations in accordance with the coupled M-pattern can be observed.  B: M-pattern velocities, determined from the 1D simulations, for different $\alpha$ and $\beta$. 
    C: The measured velocities are symmetric with respect to $\alpha$ and $\beta$ and therefore collapse on a single curve when plotted over the coupling strength $\epsilon=\sqrt{\alpha\beta}$.  Note that for small couplings the measured velocity over a longer time possibly underestimates the characteristic pattern velocity as the motion of the APS pattern is highly stochastic and seems to get pinned by the modulations due to the membrane concentration inhomogeneties. From linear stability, following the arguments of~\cite{brauns2024nonreciprocal}, we find good  quantitative agreement with the  observed front velocity of the M pattern. }
    \label{fig:velocity}
\end{figure*}

After initialization, it takes about $t_p\approx10-20 s$ for the droplets in the membrane to form and for the APS pattern $t_M\approx50-100 s$.
In 1D, the droplets are quite stable, so that after the initial formation phase the merger of two neighboring spots is rare, cf. also~\cite{BAI1994155,emmott1996coarsening,PhysRevE.71.046210} for numerical and analytical treatments of the 1D Cahn-Hilliard system.
Whereas for medium strongly coupled systems, it is quite straightforward to measure, for small coupling strengths ($\epsilon \le 0.1$), the motion of the M-pattern becomes very slow and stochastic.
From the time series it appears that the M-pattern is weakly pinned by the p-spots, which introduces slight variations to the M-subsystem, cf. figure 4A.
The interfaces of the M-spot thereby align with a p-spot over long times, before faster shifts occur that (temporarily) anchor the M-pattern in the next configuration. 
Hence, examining the system for long times ($\approx 10000s$) becomes necessary to estimate the mean pattern velocity.
The  M-pattern velocity increases and becomes more regular for larger coupling values, cf.   panel A in figure~\ref{fig:velocity}  for kymographs of the M and corresponding p-distribution for $\epsilon=0.1$ and $0.2$.
By independently varying the coupling constants $\alpha$ and $\beta$ and measuring the occurring pattern velocity $v_M$ we can quantify that $v_M$ generally increases with the coupling and is symmetric with respect to $\alpha$ and $\beta$ , cf. panel B in figure ~\ref{fig:velocity}.
Hence, for $v_M$ it does not matter which of the coupling parameters is larger or smaller, rather the coupling strength $\epsilon=\sqrt{\alpha \beta}$ determines the amplitude of $v_M$.
In particular, if either $\alpha$ or $\beta$ are zero, the M pattern is stationary, i.e., only in the fully coupled system traveling waves occur (if either $\alpha$ \emph{or} $\beta = 0$, also $\epsilon=0$).
This can easily be validated by the Jacobi Matrix of the linearized system: if either $\alpha$ or $\beta$ are zero, the Jacobian becomes singular and the eigenvalues real.
Hence, if plotted over $\epsilon$, the measured velocities collapse on a single curve, as depicted in panel C of figure 4.
This explicitly means that the here proposed coupling to a coarsening membrane is not equivalent to a static inhomogeneity in the membrane composition, and for the phenomenology of a moving pattern, both mutually coupled, dynamical systems are necessary.
We further note that a) the $p$ pattern also shifts very slightly in the opposite direction from the $M$ pattern and b) the $B$ distribution, although nearly flat, shows a variation with the $M$ pattern and a small, but noticible gradient aligned with the $M$ pattern velocity. 
More concretely, in the 1D system the $B$ distribution decreases slightly in the direction of the $M$ pattern moving.
Note that the calculated velocity for small $\epsilon$ is harder to determine, because it changes stochastically over time and therefore depends on the system realization and time interval chosen.

In order to investigate the occurrence of traveling
waves we compare the measured velocities from the 1D simulations with the estimation of the front velocity of a moving pattern via linear stability analysis as recently proposed in~\cite{brauns2024nonreciprocal}.
The authors propose that the "interface mode", i.e., the mode in the dispersion relation associated with the interface width, predicts the onset and speed of the front motion. 
Thereby, it is assumed that the underlying system is bistable, i.e., the front connects two stable, homogeneous bulk states, and the relevant mode can be determined by fitting the front profile around the inflection point $(p_{in},M_{in},B_{in})$ with a sine.
Although the linear stability analysis strictly refers to a homogeneous state (which is clearly not the case right inside the front profile), the authors argue that the analysis gives insight into the instability keeping the system from a homogeneous state as the front profile is dynamically stabilized.

Compared to that,  the situation in the present system is more involved because the $M$-pattern of the moving front is coupled to a second, inhomogeneous system $p$, representing the membrane composition.
Nevertheless, we applied the analysis with some simplifications to our case.
First, we estimate the interface mode from the uncoupled system by fitting a sine and obtain $k_{int} \approx  0.5 \mu m ^ {-1}$, which corresponds to an interface width of $\ell \approx 12.6 \mu m$ at the inflection point  $(p_{in}, M_{in}, B_{in})$. 
Due to the coupling, the inflection point is not unique anymore as the front moves over an inhomogeneous and dynamic background.
However, we assume that for small couplings, the set of possible infection points remains reasonably close to the uncoupled case.
By calculating the front velocities for different pairs of couplings $(\alpha, \beta)$ and depicting it in dependence of $\epsilon$, we obtain a very good agreement with the measured 1D velocities, cf. figure~\ref{fig:velocity} in pannel C.
Hence, we sum up that the LSA-based estimation of the front velocity of the APS pattern agrees well with the measured quantities in the 1D system.

\subsection{Coarsening Dynamics of Lipid Domains during Multiscale Pattern Formation}

\begin{figure*}[t!]
    \centering
    \includegraphics[width=0.8\linewidth]{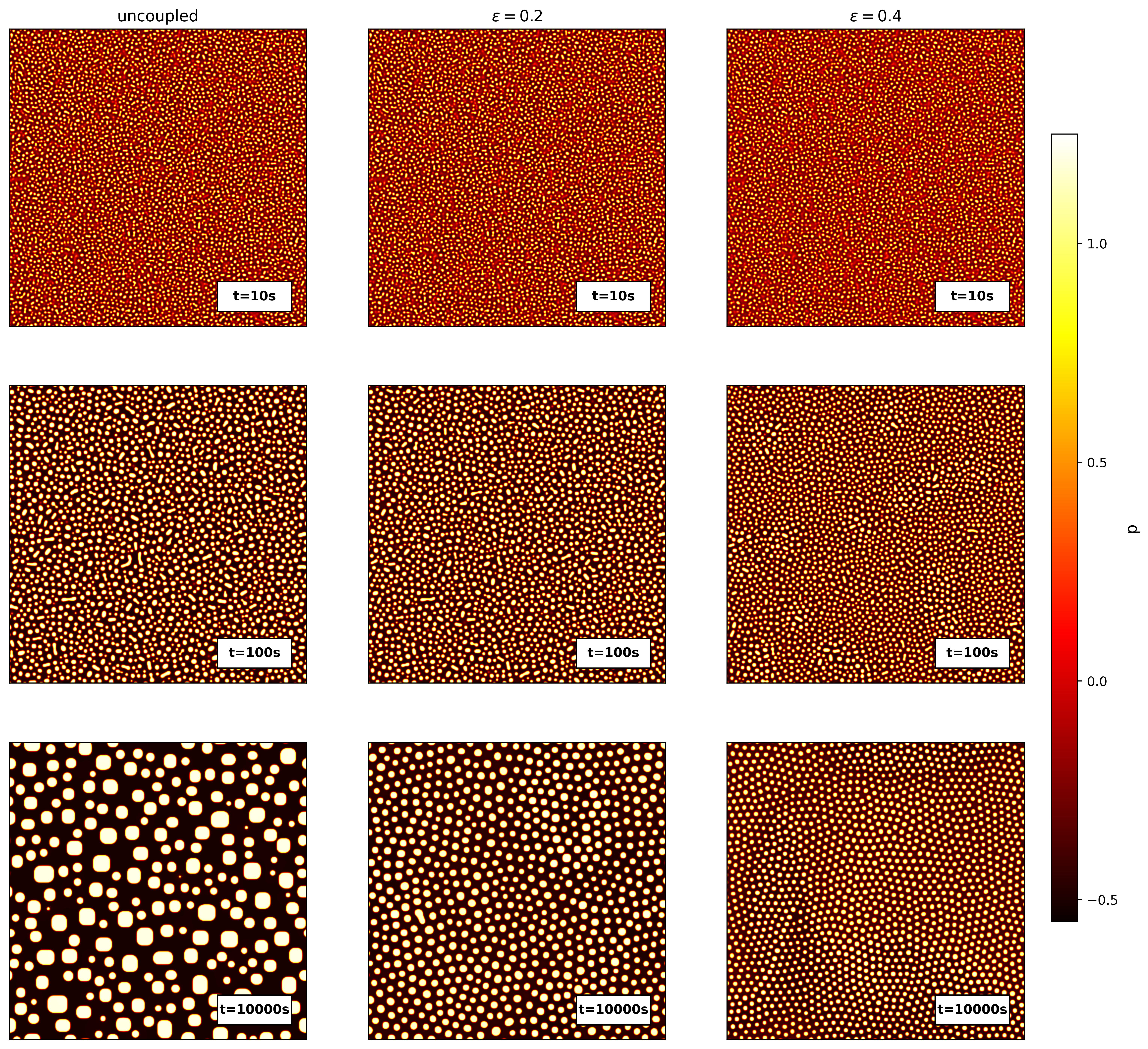}
    \caption{Snapshots of the domain evolution in the demixing Cahn-Hilliard subsystem describing the lipid membrane composition. Shown are the time points at 10s (top), 100s (middle) and 10000s (bottom) for three different coupling strengths. The uncoupled system (left) exhibits large domains (droplets) of the + phase in a background of - phase. With increasing $\epsilon$ (0.2 in the middle and 0.4 on the right), the coarsening is visibly slowed. A video showing the coarsening behavior in comparison is given in the supplementary materials as SM7 and  a systematic quantification of the slowing in figure~\ref{fig:coars}.}  
    \label{fig:coars-images}
\end{figure*}

The demixing membrane system is governed by the Cahn-Hilliard equation which is well known to exhibit coarsening behavior.
In particular, in the case of $p_0=0$, the droplets of the + phase embedded in the - phase grow with time while their number decreases.
During the study of the APS pattern velocity, we also observed that for strongly coupled systems, the droplet size of the Cahn-Hilliard system remained small and the number of droplets large.
Hence, we systematically evaluated the number and size of those droplets over time in dependence of the coupling parameters $\alpha$ and $\beta$ in order to study how  the specific coupling to the APS system alters the well-known Cahn-Hilliard coarsening.

For the systematic review, simulations over $t=10000 s$ for a domain of $67 \times 67 \mu m^2$ were performed for various coupling strengths, see figure~\ref{fig:coars-images}. 
The complete evolution of the patterns between snapshots can be seen in the movie SM7 in supplementary material. 
We calculated the mean droplet size and number of present droplets from the distribution of $p(x,y,t)$ by employing image-based edge detection, i.e., using the Canny edge detection algorithm as implemented in the python package CV2.
The number of detected contours in a frame gives the number of droplets at that time.
The mean radius was obtained by filling all contours, counting the pixels of the objects, averaging over all in the frame (giving the mean area) and taking the square root.

\begin{figure*}[t!]
    \centering
    \includegraphics[width=0.8\linewidth]{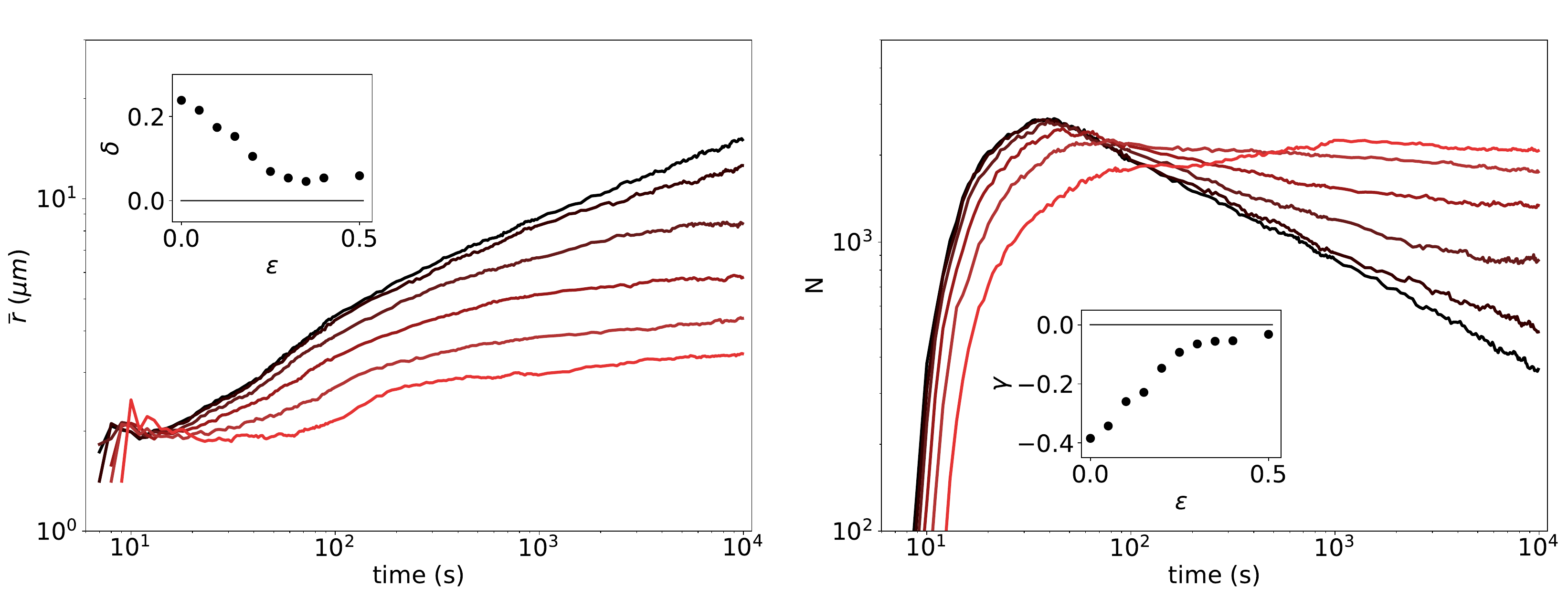}
    \caption{The coarsening of the membrane is systematically slowed by an increased coupling to the protein subsystem. Shown are the mean droplet size $\bar{r}$ (left) and number $N$ of droplets (right) in a system with $E_0=2.5$ and $p_0=0$ of  size $67 \times 67 \mu m$ over time. The coupling $\epsilon$ was increased from 0 (black) to 0.5 (red). The growth and coalescence of droplets systematically decreases with the increase in coupling strength. The inlays show the fitted slopes of the log-log plot for $t\in[10^2,10^4]$ in dependence of coupling strength $\epsilon$. Both exponents, for  $\overline{r}\propto t^{\delta}$ and $N\propto t^{\gamma}$, change nearly linearly for $\epsilon<0.25$ and remain small for larger $\epsilon$. }  
    \label{fig:coars}
\end{figure*}

The exemplary results for the symmetric case $\alpha=\beta$ are given in figure~\ref{fig:coars}  in a log-log plot highlighting the power-law behavior of the long time limit.
For both graphs, the uncoupled case is given in black, the coupling increases in increments of $\Delta \epsilon=0.1$ up to $\epsilon=0.5$ in red.
The mean radius generally increases with time, however, growth rate and mean radius both decrease with increased coupling.
By fitting the data in the long-time limit from  $t \in [1000 s,10000 s]$, we obtain the power law ($\bar{r} \propto t^{\gamma}$) with $\gamma=0.23$ for the uncoupled case, which decreases roughly linearly to $\gamma=0.03$ for the case of $\epsilon=0.25$ and remaining constant (small) for higher couplings, cf. the inlay in figure~\ref{fig:coars} for a systematic depiction.

After a fast increase in the pattern forming phase for $t<50 s$, the number of droplets $N$ decreases with time, also following a power law, as exemplified in the lower panel of fig.~\ref{fig:coars}.
Fitting $N \propto t^{\delta}$ for the same time interval reveals $\delta = -0.4$ for the uncoupled case, which roughly linearly decreases up to $\delta\approx -0.03$ for $\epsilon\approx0.25$.
For higher coupling strengths it continues to decrease but much slower.
Additionally to the symmetric case, we performed an analysis of the cases with a) constant $\alpha=0.1$, while $\beta$ varies and b) constant  $\beta=0.1$, while $\alpha$ varies and saw the same trend, but reduced in effect (data not shown here for brevity).
Hence, we find that mutually coupling the Cahn-Hilliard system with another pattern forming system, as exemplified here by equations~(1-3), slows the coarsening systematically and that \emph{both} coupling parameters contribute to the phenomenology.

\subsection{Comparison to a Simple model of Non-reciprocally Coupled Conserved Fields}

In this section, we consider a simple model of two coupled conserved field $\phi_1$ and $\phi_2$ that qualitatively reproduces the phase diagram of the more complex model for lipid-protein dynamics investigated so far in this paper. The equations describing the spatio-temporal evolution of the two fields read: 

\begin{align}
\partial \phi_1 =  -\nabla^2 ( \mu_1  \phi_1 + (\sigma  - \nu) \phi_2  + \phi_1^3  +\kappa_1   \nabla^2 \phi_1 ) && \nonumber \\ && \label{eq-phi1}\\
\partial \phi_2 =  -\nabla^2 ( \mu_2 \phi_2 + (\sigma + \nu) \phi_1 + \phi_2^3   + \kappa_2 \nabla^2 \phi_2   )&& \nonumber \\ \label{eq-phi2},
\end{align}

where $\sigma$ and $\nu$ provide a reciprocal and a non-reciprocal linear coupling between the fields $\phi_1$ and $\phi_2$, respectively. 
For the sake of simplicity we neglect nonlinear non-reciprocal coupling treated, e.g., in \cite{saha2025effervescence}, as well as cross-diffusion terms discussed, e.g., in \cite{vanag2009cross}. 
We assume that all coupling coefficients are positive quantities. 
The model can be viewed as an extension to the equations analyzed by Brauns and Marchetti in \cite{brauns2024nonreciprocal}. In this work, it was assumed that only one field $\phi_1$ is potentially unstable and consequently cubic (saturating) terms as well as diffusion was neglected, i.e., $\kappa_2 = 0$. 
Above, we already used the equation derived for wave speed of traveling domains from \cite{brauns2024nonreciprocal}. In our context, however, the conserved fields are the total protein content and the lipid composition of the membrane. 
The latter is described by a Cahn-Hilliard type dynamics already somewhat similar to the equations for the $\phi_i$ in Eqs.~(\ref{eq-phi1})-(\ref{eq-phi2}). 
Another rationale for hoping to gain insight from Eqs.~(\ref{eq-phi1})-(\ref{eq-phi2}) is given by the findings by Bergmann et al. \cite{bergmann2018active,bergmann2019system} that a reaction-diffusion model coupling membrane and bulk concentration of proteins and exhibiting active separation yields also a Cahn-Hilliard equation as amplitude equation near the onset of the instability of the spatially homogeneous steady state. 
Since in our model the spatially homogeneous steady state of both fields can become unstable, we need to include cubic terms in both Eq.~(\ref{eq-phi1}) and Eq.(~\ref{eq-phi2}). 
Further, we assume that the effective diffusivities $\kappa_1$ and $\kappa_2$ may be quite different, leading to instabilities of the uncoupled systems at very different wave numbers. 
This fact was shown in the context of protein-lipid systems already a while ago \cite{john2005travelling,john2005alternative} and is visible also in the dispersion curves of the full model described above. 
In general, protein dynamics provides spatial coupling at considerably larger scale than the lipid dynamics in the membrane. 
If we assume $\phi_1$ stands for the conserved protein content and $\phi_2$ stands for the lipid composition of the membrane,  one needs to consider $\kappa_1 >> \kappa_2$.

\begin{figure*}[t!]
    \centering
    \includegraphics[width=0.75\linewidth]{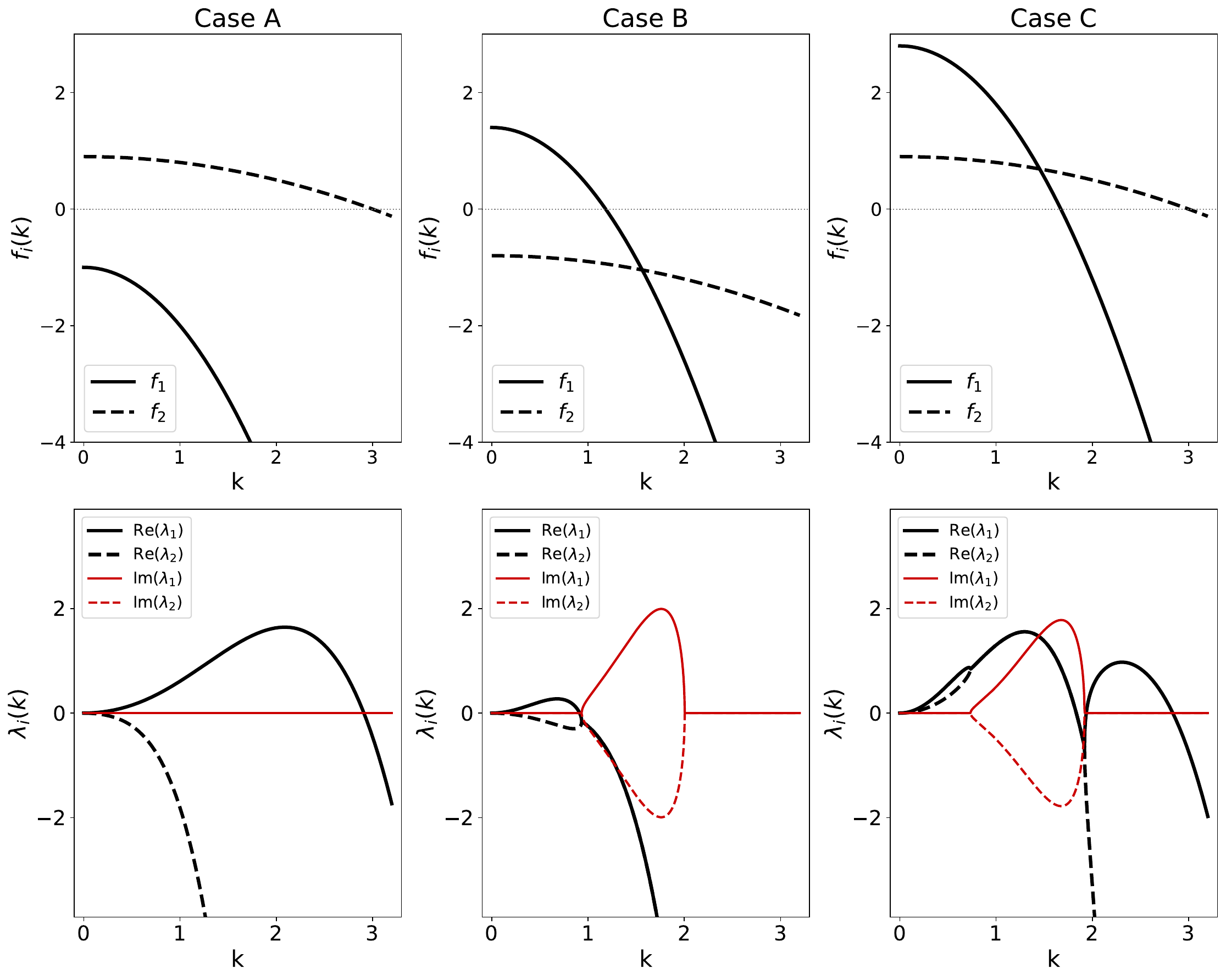}
    \caption{
        Solutions of the minimal model for non-reciprocal coupling. Depicted are the functions $f_1$ and $f_2$ (top) and the corresponding eigenvalues (bottom) for different system parameters. 
        \textbf{Case A:} For $\mu_1=-1$, $\mu_2=0.9$, $\kappa_1=1$ and $\kappa_2=0.1$ in the minimal system, the $f_i$ do not intersect and are both monotonously decreasing with $k$. The corresponding eigenvalues are real and only one exhibits a positive maximum at $k\approx2$, whereas the second eigenvalue is monotonously decreasing with $k$ from $\lambda_2(0)=0$. This corresponds to a single static pattern. 
        \textbf{Case B:} For $\mu_1=1.4$, $\mu_2=-0.8$, $\kappa_1=1$ and $\kappa_2=0.1$, both $f_i$ intersect at $k\approx1.5$ at a negative function value. The eigenvalues become complex in an intermediate regime ($1<k<2$), where a pair of complex conjugate eigenvalues emerges. The real part of $\lambda_1$ is positive for $k<0$, whereas $\operatorname{Re}(\lambda_2)$ is always $\leq 0$. The maximum of $\operatorname{Re}(\lambda_1)$ corresponds to the typical wavenumber of a single static pattern, as in this case all unstable modes remain real. 
        \textbf{Case C:} For $\mu_1=2.8$, $\mu_2=0.9$, $\kappa_1=1$ and $\kappa_2=0.1$, the functions $f_i$ intersect at a positive value at $k\approx1.5$. Here, we observe two positive maxima of the real parts of the eigenvalues, corresponding to two separate patterns of different size. In the intermediate region, where the $\lambda_i$ are complex, also positive real parts coexist, corresponding to one of the patterns exhibiting an oscillatory instability. Videos of the explicit model simulations for cases A-C and case C with reciprocal coupling are attached in the supplementary material as SM 8a, 8b, 8c and 8c2.
    }
    \label{fig:schematicEVs}
\end{figure*}

A linear stability analysis of Eqs.~(\ref{eq-phi1})-(\ref{eq-phi2}) around the steady state $ (\phi_1, \phi_2) = \bf{\phi_0} = (0, 0)$ assuming perturbations $ \propto \exp (ikx + \lambda t)$ and setting $\mu_1 =  \mu_2 = \mu$  for convenience is straightforward and produces the following dispersion curves for the eigenvalues $\lambda_i$ 

\begin{align} 
\lambda_{1,2} (k) = k^2 \left( (f_1 + f_2)/2 \pm \sqrt{ (f_1 - f_2)^2/4 + \sigma^2 - \nu^2}\right), 
\label{dispersion_phi_i}
\end{align}

where $f_1 = \mu_1 - \kappa_1 k^2$ and $f_2 = \mu_2 - \kappa_2 k^2$ are the eigenvalues for the uncoupled fields.
 One can see immediately that in order to have oscillatory eigenvalues the condition 
\begin{align}
\nu^2 > \sigma^2 +  (f_1 - f_2)^2/4.
\label{cond-osc-inst}
\end{align}
has to be fulfilled. 
This shows that (a) the non-reciprocal coupling has to be strong enough and dominant over the reciprocal coupling and (b) that the values of the $\mu_i$ and $k$ where $f_1$ and $f_2$ become equal are most likely to have oscillatory eigenvalues. 
If one assumes that $\phi_1$ is the unstable and $\phi_2$ the stable field ($\mu_1 > 0 > \mu_2)$ the conditions for a long-wavelength oscillatory instability (type $II_O$ after \cite{cross1993pattern}) becomes $\mu_1 = - \mu_2$ and $\nu^2 >  \mu_1^2 + \sigma^2$.  
If  the coupling between the fields $\phi_1$ is weak {\it i. e.} $|\sigma^2 - \nu^2| << \mu_1 - \mu_2$, the condition for oscillatory eigenvalues, see Eq.~(\ref{cond-osc-inst}), is not fullfilled at zero wavenumber and the instability becomes of type $II_S$ after \cite{cross1993pattern}. 
The eigenvalues are then $\lambda_1 \approx k^2 (f_1 + (\sigma^2 - \nu k^2))$ and $\lambda_2\approx f_2 -(\sigma^2 - \nu k^2))$.  If non-reciprocal coupling dominates as for the full model above (otherwise no oscillations or traveling waves would be possible), the condition for the instability at $k = 0$ writes $\mu_1 \approx - \mu_2 + \nu^2 - \sigma^2 $, i. e., the coupling is stabilizing. 
If we suppose that $\phi_1$ describes the lipid composition, one needs to set $\kappa_1 << \kappa_2$ and it immediately follows that if the primary instability is of type $II_S$ no oscillatory eigenvalues will occur for any perturbation wavenumber $k$. 
This is indeed what we found from the LSA of the full system, cf. section~\ref{LSA}. 
For the opposite case where $\phi_1$ stands for the lipids, one can also easily find that if the coupling between the fields increase, oscillatory eigenvalues occur first at larger values of $k$ quite similar to the scenario discussed in \cite{brauns2024nonreciprocal}. 
Altogether, all dispersion curves found for the full model above in Section~\ref{LSA} can also be reproduced in the simplified model as is demonstrated in figure~\ref{fig:schematicEVs}.
If we use an analogy to the reasoning for classical activator-inhibitor models, we may say traveling waves occur if the activator diffuses fast - this is the case where the protein concentrations provide the instability.
On the other hand, stationary patterns are expected when the inhibitor diffuses fast - this is the case where the lipid composition can become unstable.
Hence, the comparison to the simple model here shows that the full model has many of characteristic properties of non-reciprocal pattern formation (to borrow the term from \cite{brauns2024nonreciprocal}). 
A comparison of the pattern dynamics in the full model described above (see movie SM6 in supplementary materials) with the pattern dynamics in the reduced model described in this section (see movies SM8a, 8b, 8c-1, 8c-2) shows that the patterns in the simple model are similar for both coupled conserved fields and fail to reproduce the dynamics in the region of multiscale pattern formation. 
This indicates that the dynamics in the full model far away from the linear instabilities is not described by the simplified model, which is closer in nature to the amplitude equation valid near onset of oscillatory separation \cite{frohoff2023nonreciprocal,greve2024amplitude}. 
Altogether, the linear stability analysis simplified model explain well the asymmetry in the instabilities of the homogeneous protein (oscillatory) and lipid concentrations (stationary). The simple equations in this section already show that for coupled conserved fields with long-wavelength, phase separating instabilities on largely different lengthscales, traveling domains are only possible for the species that exhibits larger characteristic length scales as observed in the full model above as well as in the detailed model of MARCKS protein dynamics a while ago \cite{john2005travelling}.

\section{Conclusions}

We studied a biophysical model for cell polarization via active phase separation due to the binding and unbinding of a freely diffusing, intracellular protein to the surrounding cell membrane.
For homogeneous membrane composition this frequently studied model~\cite{mori2008wave}exhibits a singular, stationary spot pattern spanning the whole cell.
Here, we coupled it to a model of a cell membrane consisting of two different lipid species that are governed by the Cahn-Hilliard equation and hence undergo phase separation, typically leading to  a pattern of  smaller spots that very slowly coarsen over time. 
By coupling the dynamics of both systems, we observed novel behavior including traveling protein domains and multiscale patterns, i.e., lipid and protein patterns at substantially different length scales. 
The onset of motion of the bigger protein pattern is related to  a long-wavelength oscillatory instability (type II-O instability).
The coupling to the protein subsystem further leads to a slowing down or even arrest of the  coarsening of the lipid domains in the membrane.

Moreover, we observe that the static protein pattern starts to move at a finite coupling strength $\epsilon \gtrsim 0.1$.
Just above this threshold, this motion is highly stochastic, i.e., exhibiting long times of static $M$ pattern with fast, intermittent motion that change direction of motion frequently.
As $\epsilon$ grows, the pattern velocity becomes more regular and increases monotonously.
We show that the front velocity at sufficiently strong coupling can be predicted from the linear stability analysis of the full system by studying the so-called "interface mode", i.e., the wave number associated with the interface width of the pattern in question, following an argument given in ~\cite{brauns2024nonreciprocal}.
The coarsening of the membrane composition is slowed down systematically and eventually suppressed with increasing coupling strength $\epsilon$, which can be quantified by fitting power-law behavior to the long-time tails of the average spot radius and the number of spots over time.
From the exponents' behavior with increasing coupling strength $\epsilon$ we learn that the decrease of the coarsening behavior  is first roughly linear with $\epsilon$ for $\epsilon \lesssim 0.25$.
For larger values of $\epsilon$, the coarsening scaling is very slow and remains roughly constant, cf. fig.~\ref{fig:coars} the two inlays.

The studied APS system via membrane interaction represents a simplified model to describe the polarity transition in biological cells.
There, the single spot of protein-rich phase on the membrane leads to a so-called prepatterning that initializes more permanent intracellular changes leading to a persistent polarization of the cell and with that, enables crucial cell function in morphogenesis and differentiation~\cite{drubin1996origins}.
Hence, in this context the observed onset of motion of the protein pattern exemplifies that the coupling of multiple systems in order to describe more complex behaviors may lead to phenomenological changes of the different subsystems that may also  impede their (primary) function.
I.e., a moving protein pattern is opposed to the function of establishing a persistent signal or symmetry break for cell polarity.
However, we here presented a minimal model and the underlying biological system is often more complex.
Nevertheless, the simple model for protein-lipid dynamics at membranes shows that the dynamics of lipids, which typically lead to 
patterns of submicrometer length scale, can change the qualitative behavior of the protein domains on much larger length scale. 
Naively, one may argue that the proteins "see" only the average of the lipid dynamics and therefore no explicit modeling of the lipid dynamics is
needed, the model here clearly shows that the coupling of both dynamics immediately leads to traveling protein dynamics even in absence of a second 
protein dynamics or of actin dynamics. 
While both lipid domains as well as protein patterns are frequently found in a variety of experimental cellular (and sometimes in-vitro) systems, 
a convincing experiment illustrating the role of the lipid-protein patterns is still missing despite some preliminary findings showing oscillatory transients \cite{alonso2011oscillations}.

Furthermore, we have simplified the model to obtain the reduced form presented in Eqs.~(\ref{eq-phi1}--\ref{eq-phi2}). 
To analyze and interpret the resulting dispersion curves of the full model, we adapted a previously proposed minimal model for non-reciprocal interactions \cite{brauns2024nonreciprocal} to our specific case. 
Our analysis indicates that the simplified two-variable  model reproduces the characteristic features of the linear stability analysis of the full three variable model  discussed in this paper. 
In the simple model it can also be  shown analytically that oscillatory instabilities do occur for the species with larger interface tension, also corresponding to a higher characteristic initial phase separation length scale. 
The simplified model further provides a generalization of the frequently 
studied non-reciprocal Cahn-Hilliard equations described in the introduction and is a promising candidate to study more qualitative aspects of 
non-reciprocal pattern formation.  
The nonlinear dynamics in the two-variable phenomenological model is, however, more restricted and does not fully reproduce the  observations of the full three-variable model given by eq. \ref{eq_sys1}-\ref{eq_sys3}.  

On a more abstract level we show how, through the interaction of multiple subsystems, complexity in, e.g., biological and softmatter systems can emerge.
When well-studied minimal models exist that by nature capture only a small part of a more complex reality, it is a straight-forward approach to investigate the interplay of different systems by introducing coupling terms.
Starting from the assumption that single models are well justified in their modular nature, the coupling-mediated behavior can then be a systematic tool to explore emerging complex behavior in the low-coupling limit and beyond.

\bibliography{biblio}
\bibliographystyle{unsrt}

\end{document}

\section*{Supplementary Material}

\subsection*{SM1}

\begin{figure}[h!]
    \centering
    \begin{subfigure}[b]{0.52\textwidth}
        \centering
        \includegraphics[width=\textwidth]{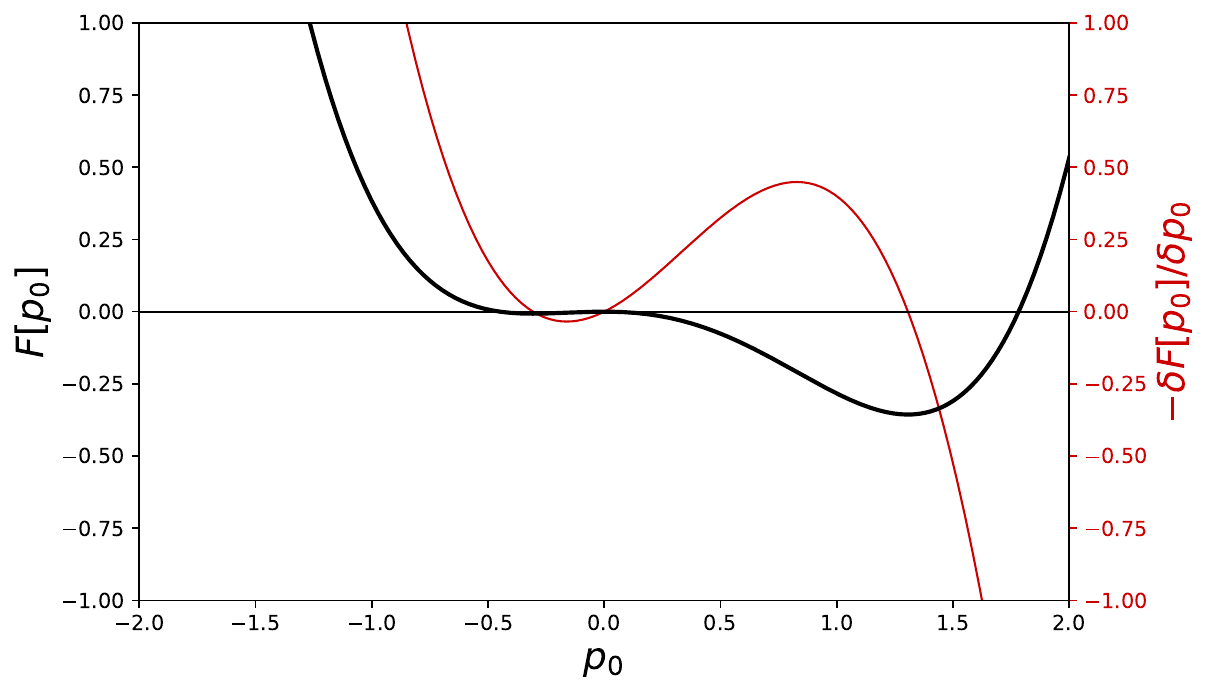}
    \end{subfigure}
    \hfill
    \begin{subfigure}[b]{0.46\textwidth}
        \centering
        \includegraphics[width=\textwidth]{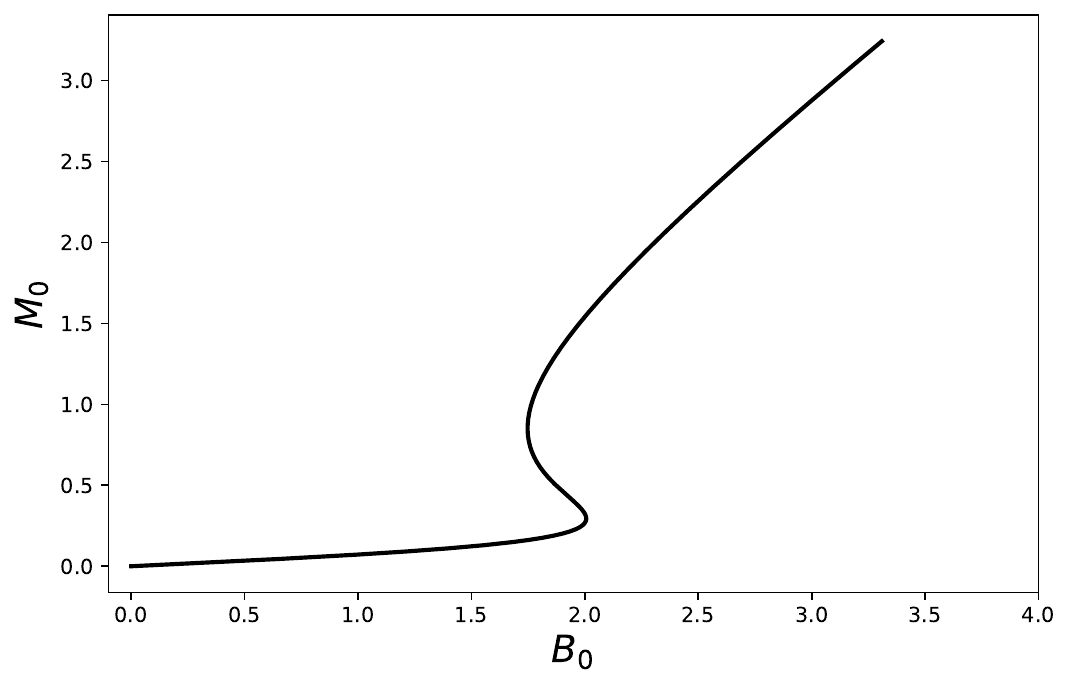}
    \end{subfigure}
    \caption*{\textbf{SM1:} Steady states of the subsystems in the absence of coupling. Left: The potential of the Cahn-Hilliard model (black) for the chosen parameters ($\tau=0.4,\, \Psi=0.1$), cf. equation (3) in the main text. As well as the variational with respect to $p_0$, $\frac{\delta F[p_0]}{\delta p_0}$ (red) which determines the time evolution of the system. Note that the two stable states at -0.48 and 1.48  correspond to the two phases in the simulation. Right: The reaction nullkline of the M-B protein subsystem; given by the balance condition of the binding $(f_a)$ and unbinding rates $(f_d)$. In the absence of gradients, it follows then that $d/dt M = d/dt B = 0$. Note the bistable region around $B_0\approx [1.75,2]$ with an upper and a lower stable level of $M(B)$ and a negative slope $dM/dB<0$ that represents an unstable state. }
    \label{fig:SM1}
\end{figure}

\subsection*{SM2}

Video of the linear stability of the coupled system with $\epsilon=0.3$. The increase in $p_0$ with constant $E_0=1.8$ corresponds to the case A in figure 3 of the main manuscript. A single eigenvalue becomes positive for the critical wavenumber $k_c$ before developing a widening band of unstable modes. As the imaginary part of the eigenvalues remains zero, it is a static instability.

\subsection*{SM3}

Video of the linear stability of the coupled system with $\epsilon=0.3$. The increase in $E_0$ with constant $p_0=-0.25$ corresponding to the case B in figure 3 of the main manuscript. The imaginary part of the eigenvalue becomes nonzero as the real part becomes positive for $k_c$. This marks an oscillatory instability.

\subsection*{SM4}

Video of the linear stability of the coupled system by increasing the symmetric coupling from $\epsilon=0$  to $0.2$ for $p_0=0.3$ and $E_0=2.3$, corresponding to the point C in the phase diagram of fig. 3 in the main manuscript.
The smaller pattern of Cahn-Hilliard subsystem is unstable (maximum at large $k_{CH}$, here out of frame) and the APS pattern becomes unstable for $\epsilon>0$ at very small $k_{APS}\approx0$. The APS-related maximum of the real part quickly shifts towards higher k values together with a nonzero imaginary part at slightly higher k - constantly placed between the maxima $k_{CH}$ and $k_{APS}$.

\subsection*{SM5}

Video of the linear stability of the coupled system similar to AM6, but for the initial condition for the 1D and 2D simulations of the system, i.e., at $p_0=0$ and $E_0=2.5$.
Both pattern are already unstable (for $\epsilon=0$) and with $\epsilon>0$ an intermediate region win nonzero imaginary part of the eigenvalue develops.

\subsection*{SM6}

Video of the coupled system in 2D over 1000s. The parameters are: $\epsilon=0.3$, $E_0=2.5$, $p_0=0$ with a system size of $33.5\,\times\,33.5\mu m$. p and M concentrations are depicted by a color gradient - from deep blue over green to yellow - in the constant ranges [-0.55,1.25] and [0.1,1.7], respectively.

\subsection*{SM7}

Video of the coarsening of the lipid subsystem (p only) for three different couplings: $\epsilon=0$, $0.2$ and $0.4$.
The color gradient goes from -0.51 to 1.15 for a system size of $33.5\,\times\,33.5\mu m$. The time steps $\Delta t$ increase from 1s (for $t < 100s$) to 10s (for $100s < t < 1000s$) and to 100s (for $1000s < t < 10000s$). The larger moving phases with less contrast are the influence of the coupled APS subsystem upon the coarsening lipids.

\subsection*{SM8a}

Video of the minimal model, cf. equations 4 and 5, for case A in figure 7 of the main text.
The parameters are $\mu_1=-1$, $\mu_2=0.9$, $\kappa=1$ and $\nu=0.1$ for a non-reciprocal coupling of $-0.5$.
In this case, the linear stability analysis yields one real eigenvalue with a broad unstable band of k-numbers, whereas the second eigenvalue is stable for all k.
Shown are the $\Phi_1$ (left) and $\Phi_2$ field (right) for $100s$ on a 2D domain with $100\mu m$ length and periodic boundary conditions.

\subsection*{SM8b}

Video of the minimal model, cf. equations 4 and 5, for case B in figure 7 of the main text.
The parameters are $\mu_1=1.4$, $\mu_2=-0.8$, $\kappa=1$ and $\nu=0.1$ for a non-reciprocal coupling of $-0.5$.
In this case, the linear stability analysis yields one complex eigenvalue with an unstable band at low k-numbers, that is followed by an interval of non-zero imaginary part but negative real part.
The second eigenvalue is stable for all k and shares the interval of non-zero imaginary part as a pair of complex conjugate values.
Shown are the $\Phi_1$ (left) and $\Phi_2$ field (right) for $100s$ on a 2D domain with $100\mu m$ length and periodic boundary conditions.

\subsection*{SM8c}

Video of the minimal model, cf. equations 4 and 5, for case C in figure 7 of the main text.
The parameters are $\mu_1=2.8$, $\mu_2=0.9$, $\kappa=1$ and $\nu=0.1$ for a non-reciprocal coupling of $-0.5$.
In this case, the linear stability analysis yields two complex eigenvalues with two separate unstable bands of k-numbers.
One of the bands also exhibits non-zero imaginary part, indicating the existence of traveling waves and hence, a moving pattern.
In the interval of non-zero imaginary part, again both eigenvalues form a pair of complex conjugate values.
Shown are the $\Phi_1$ (left) and $\Phi_2$ field (right) for $100s$ on a 2D domain with $100\mu m$ length and periodic boundary conditions.

\subsection*{SM8c-2}

Video of the minimal model, cf. equations 4 and 5, for case C but for a reciprocal coupling of $0.5$ (not shown in the main text).
Here, the eigenvalues remain real, leading to a much different phenomenology, i.e., the pattern exhibits no actively moving fronts as in the cases A and B.
Shown are the $\Phi_1$ (left) and $\Phi_2$ field (right) for $100s$ on a 2D domain with $100\mu m$ length and periodic boundary conditions.